\documentclass[twocolumn,aps,prc,showpacs,superscriptaddress,preprintnumbers,floatfix,nofootinbib]{revtex4}
\usepackage{epsfig,graphics}
\usepackage{graphicx}
\usepackage{dcolumn}
\usepackage{bm}
\usepackage{amsmath}
\usepackage[usenames]{color}
\usepackage{ulem} 

\begin{document}


\title{Pygmy and Giant Dipole Resonances by Coulomb Excitation using a Quantum Molecular Dynamics  model}

\author{ C. Tao}
\affiliation{Shanghai Institute of Applied Physics, Chinese
Academy of Sciences, Shanghai 201800, China}
\affiliation{
University of the Chinese Academy of Sciences, Beijing 100080, China}
\author{ Y. G. Ma}
\thanks{Author to whom all correspondence should be addressed. Email: ygma@sinap.ac.cn}
\affiliation{Shanghai Institute of Applied Physics, Chinese
Academy of Sciences, Shanghai 201800, China}
\author{ G. Q. Zhang}
\affiliation{Shanghai Institute of Applied Physics, Chinese
Academy of Sciences, Shanghai 201800, China}
\author{ X. G. Cao}
\affiliation{Shanghai Institute of Applied Physics, Chinese
Academy of Sciences, Shanghai 201800, China}
\author{ D. Q. Fang}
\affiliation{Shanghai Institute of Applied Physics, Chinese
Academy of Sciences, Shanghai 201800, China}
\author{ H. W. Wang}
\affiliation{Shanghai Institute of Applied Physics, Chinese
Academy of Sciences, Shanghai 201800, China}

\date{ \today}

\begin{abstract}

Pygmy and Giant Dipole Resonance (PDR and GDR) in Ni isotopes
have been investigated by Coulomb excitation in the framework of
the Isospin-dependent Quantum Molecular Dynamics model (IQMD). The spectra of $\gamma$
rays are calculated and the peak energy, the strength and Full
Width at Half Maximum (FWHM) of GDR and PDR have been extracted.
Their sensitivities  to nuclear equation of state, especially to its symmetry
energy term are also explored. By a comparison with the other mean-field
calculations, we obtain the reasonable values for symmetry energy and its slope parameter at saturation, which gives an important constrain for IQMD model.
In addition, we also
studied the neutron excess dependence  of GDR and PDR parameters for
Ni isotopes and found that the energy-weighted sum rule (EWSR)
$PDR_{m_1}/GDR_{m_1}\%$ increases linearly with the neutron excess.

\end{abstract}


\maketitle

\section{Introduction}

Intermediate energy heavy-ion collisons provide a chance to study reaction machanism and 
the propertis of hot nuclei, such as fragmentation and liquid gas phase transition etc \cite{Bor,Adv,Nat,Ma,LiSX}. In a relative low excitation
region of nuclei, dipole resonance has attracted much attention experimentally and
theoretically in the past few decades ~\cite{PDRReview1,PDRReview2}.
Two kind of resonances have been evidenced, one is Pygmy Dipole Resonance (PDR) 
and another is  Giant Dipole Resonance
(GDR).  In contrast to GDR, which can be considered as the oscillation between
non-deformed, incompressible proton and neutron spheres, PDR can
be considered as the oscillation between the weakly-bound neutron
skin and the isospin neutral proton-neutron core.  From a
theoretical point of view, the presence of this low-lying strength
PDR is predicted by almost all microscopic models, ranging from
Hartree-Fock plus random phase approximation (RPA) with Skyrme
interactions to relativistic Hartree-Bogoliubov plus relativistic
quasiparticle RPA \cite{PDRReview2}.  Experiments of
PDR~\cite{PDREXP1,PDREXP2,PDREXP3,nidat,sndat,PDREXP4,PDREXP5} in
neutron-rich nuclei and microscopic models which have been applied
to investigate
PDR~\cite{MAZY,PDRCAL2,PDRCAL3,PDRCAL4,PDRCAL5,PDRCAL6,Liu}
have shown that the PDR could have a pronounced relationship with
neutron-capture rates in the $\gamma$-process, nucleosynthesis,
the radiative neutron-capture cross section on neutron-rich
nuclei, and the photodisintegration of ultra-high energy cosmic
rays. The properties of the PDR in stable nuclei have been studied
extensively for different neutron and proton shell closures with
the ($\gamma$,$\gamma'$) reactions. However, studies for neutron-rich
isotopes, where the PDR should be enhanced, were still less, especially in the
framework of molecular-dynamics type model.
Recently,  the influence of  symmetry energy on PDR has been also
checked  \cite{MaZY,Baran}. For example, in Ref.~\cite{MaZY}, they studied  isovector giant and pygmy dipole resonances in even-even Ni isotopes within the framework
of a fully consistent relativistic random-phase approximation built on the relativistic mean field ground state and found that the centroid energy of the isovector pygmy resonance is insensitive to the density
dependence of the symmetry energy.  
Ref.\cite{Baran} used a semiclassical Landau-Vlasov approach to investigate the energy centroid associated with the PDR and found that PDR is insensitive to the symmetry energy term of EOS. Considering symmetry energy is one of interesting topics in current heavy-ion physics community, we would also like to check this sensitivity in the present work.

In this work, we  apply the Isospin-dependent Quantum Molecular Dynamics model (IQMD) to investigate the properties of PDR. In our previous work,
the same model has been successfully applied into the GDR calculation.
Here we plan to extend the model to calculate PDR in the same framework. The
molecular dynamics model is a kind of Monte-Carlo transport model, which
has been extensively applied in heavy-ion collision dynamics. In the model,
the physical parameters can be controlled and then their respective
effects on PDR and GDR can be addressed.

The paper is organized as follows. Sec. II gives a brief introduction of IQMD model as well as the formalism to calculation GDR and PDR in the IQMD framework.
Results and discussions are presented in Sec. III where different parameter dependencies of GDR
and PDR are shown. In particular, the equation of state, symmetry energy and isospin effects are presented and discussed. Finally the summary is given in Sec. IV.

\section{Model and formalism }

The isospin-dependent quantum molecular dynamics (IQMD) model is based
on QMD model  \cite{QMDAichelin}. In IQMD model, the mean field is
given by: $U(\rho)=U^{Sky}+U^{Coul}+U^{Yuk}+U^{sym}+U^{MDI}$,
where $U^{Sky}$, $U^{Coul}$, $U^{Yuk}$, $U^{sym}$ and $U^{MDI}$ is
the Skyrme potential, Coulomb potential, Yukawa potentia,
symmetry potential interaction, and  the momentum dependent
interaction (MDI), respectively. The Skyrme potential can be presented
as follows:
\begin{equation}
U^{Sky}=\alpha (\rho /\rho _0)+\beta (\rho /\rho _0)^\gamma,
\label{eq01}
\end{equation}
where $\rho _0=0.16/fm^{3}$ (the saturation nuclear density) and
$\rho$ is the nuclear density. In the equation, different
[$\alpha$, $\beta$, $\gamma$] represents different  kinds of
equations of state (EOS). The parameters $\alpha$, $\beta$ and
$\gamma$ are given in Table~\ref{table1}. In the Table, S(M)
represents the soft EOS (with MDI),
with an incompressibility of $K$=200 MeV, while H(M)
for the hard EOS(with MDI), with an incompressibility of $K$=380 MeV.
\begin{table}[htbp]
\caption{The parameters $\alpha$, $\beta$ and $\gamma$ for the different EOS}
\label{table1}
\centering
\begin{tabular}{p{42pt}p{42pt}p{42pt}p{42pt}p{42pt}}
\hline
\hline
EOS  &$K$ & $\alpha$ & $\beta$ & $\gamma$ \\
 & (MeV) & (MeV) & (MeV) & (MeV) \\
\hline
S & 200 & -356 & 303 & 7/6 \\
SM & 200 & -390.1 & 320.3 & 1.14 \\
H & 380 & -124 & 70.5 & 2 \\
HM & 380 & -129.2 & 59.4 & 2.09 \\
\hline
\hline
\end{tabular}
\end{table}

$U^{Coul}$, $U^{Yuk}$, $U^{sym}$ and $U^{MDI}$ can be expressed as follows, respectively:
\begin{equation}
U^{Coul}=\frac{e^2}{4}\sum_{i\neq j}\frac{1}{(4\pi L)^{(3/2)}}exp[-\frac{\lvert  r_i-r_j \rvert^2}{4L}].
\label{eq02}
\end{equation}
Here, $r_{ij}=\lvert \textbf{r}_i-\textbf{r}_j\rvert$ represents
the relative  distance of two nucleons, and the $L$ is the so-called Gaussian
wave-packet width for nucleons. For simplicity,  here we choose a constant wave-packet width, i.e.
$L$= 2.16 fm$^2$. We noted that there was some discussions on the affect of the width on the dynamical results, eg.  flow, multifragmentation, pion and kaon production etc \cite{Puri2}. However, we are treating the Coulomb excitation of the projectile in which the dynamical effect is relatively smaller. 


\begin{equation}
\begin{array}{l}
U^{Yuk}=(V_y/2)\sum_{i\neq j}\frac{1}{r_{ij}}exp(Lm^2) \\
\hspace{3.7em}
\times [exp(mr_{ij})erfc(\sqrt{L}m-r_{ij}/\sqrt{4L}) \\
\hspace{3.7em}
-exp(mr_{ij})erfc(\sqrt{L}m+r_{ij}/\sqrt{4L})],
\label{eq03}
\end{array}
\end{equation}
where $Vy$=0.0024 GeV and $m=0.83$.

\begin{equation}
U^{sym}=\frac{C_{sym}}{2\rho _0}\sum_{i\neq j}\tau _{iz}\tau _{jz}\frac{1}{(4\pi L)^{3/2}}exp[-\frac{(\textbf{r}_i-\textbf{r}_j)^2}{4L}],
\label{eq04}
\end{equation}
where $C_{sym}$ is the symmetry energy coefficient, $\tau _z$ is the
$z$th  component of the isospin degree of freedom for the nucleon, which equals 1 or -1 for neutron or proton,
respectively.

\begin{equation}
U^{MDI}=\delta ln^2[\varepsilon (\frac{\rho}{\rho _0})^2+1](\frac{\rho}{\rho _0}),
\label{eq05}
\end{equation}
Here, $\delta$ =1.57 MeV and $\varepsilon$ =500 $c^2/$GeV$^2$.
More details of the QMD model can be found in~\cite{QMDAichelin,Hartnack1989,Hartnack1998,Ma2006,Puri}.

The dipole moment of GDR in coordinator space [$DR_{GDR}(t)$] and
momentum space [$DK_{GDR}(t)$] is respectively defined as
follows~\cite{BaranNPA679,Wu}:
\begin{eqnarray}
DR_{GDR}(t)=\frac{NZ}{A}[R_{Z}(t)-R_{N}(t)],\\
\label{eq1}
DK_{GDR}(t)=\frac{NZ}{A\hbar}[\frac{P_{Z}(t)}{Z}-\frac{P_{N}(t)}{N}],
\label{eq2}
\end{eqnarray}
where  $R_{Z}(t)$ and $R_{N}(t)$ are the center of mass of protons
and neutrons in  coordinator space, respectively; $P_{Z}(t)$ and
$P_{N}(t)$ are the center of mass of protons and neutrons in
momentum space, respectively.

By the Fourier transformation of the second derivative of $DR_{GDR}(t)$ in respect to time, i.e.
\begin{equation}
DR''(\omega)=\int_{t_0}^{t_{max}}DR''(t)e^{i\omega t}dt,
\label{eq3}
\end{equation}
we can get the the $\gamma$ ray emission probability for energy $E_\gamma = \hbar \omega$  as follow:
\begin{equation}
\frac{dP}{dE_{\gamma}}=\frac{2e^2}{3\pi \hbar c^3E_{\gamma}}\lvert DR''(\omega)\rvert ^2.
\label{eq4}
\end{equation}

Similarly, we can write the dipole moment of PDR in coordinator space and momentum space as follows:
\begin{eqnarray}
DR_{PDR}(t)=\frac{N_vZ}{A}[R_{C}(t)-R_{N_v}(t)],\\
\label{eq5}
DK_{PDR}(t)=\frac{N_vZ}{A\hbar}[\frac{P_{C}(t)}{2Z}-\frac{P_{N_v}(t)}{N_v}],
\label{eq6}
\end{eqnarray}
where  $R_{C}(t)$ and $R_{N_v}(t)$ are the center of mass of
isospin-symmetric core and  valence neutrons in coordinator space,
respectively; $P_{Z}(t)$ and $P_{N}(t)$ are the center of mass of
isospin-symmetric core and valence neutrons in momentum space,
respectively. Then we can also get the $\gamma$ ray emission
probability of PDR. Here, we choose neutrons with the
farthest distances from the CM of all nucleons as
the valence neutrons in the initial state.

The fraction of the energy-weighted sum rule (EWSR) contained in the PDR
relative to that  located in the GDR region can be written as
follows
\begin{equation}
m_1\%=\frac{PDR_{m_1}}{GDR_{m_1}}\times 100\%,
\label{eq7}
\end{equation}
where
$GDR_{m_1}$ and $PDR_{m_1}$  is EWSR of GDR and PDR, respectively:
\begin{eqnarray} 
GDR_{m_1}=\sum_{E_1}^{E_2}(\frac{dP}{dE_{\gamma}})_{GDR}\Delta E\times E,\\
\label{eq8}
PDR_{m_1}=\sum_{E_1'}^{E_2'}(\frac{dP}{dE_{\gamma}})_{PDR}\Delta E\times E.
\label{eq9}
\end{eqnarray}

\section{Results and discussions}

\begin{figure}[htbp]
\resizebox{8.6cm}{!}{\includegraphics{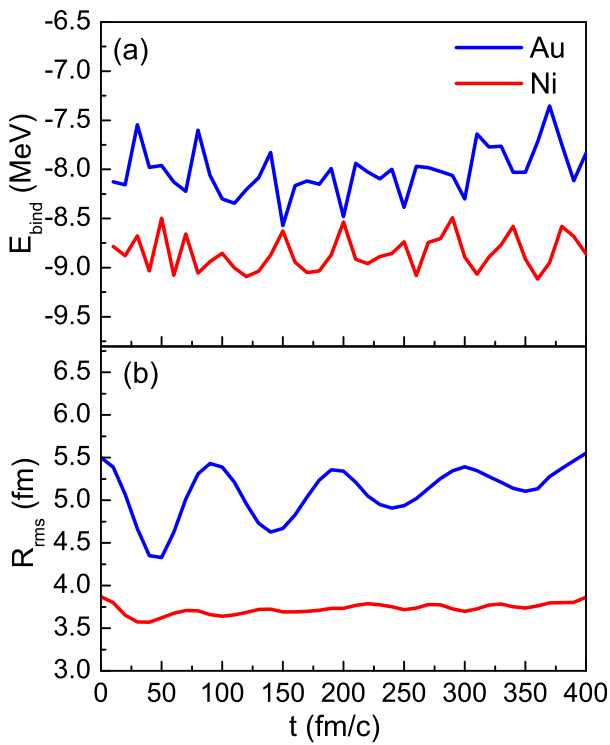}}
\vspace{-2cm}
\caption{(Color online) Stability check for the projectile ($^{68}$Ni) and target ($^{197}$Au).
(a) time evolution of the binding energies; (b) time evolution of rms radii. } \label{stability}
\end{figure}

Now we move on the calculations and discussions. As a first step of this work, we need to check the stability of the nuclei \cite{Ma2006}. Here we select
$^{68}$Ni+$^{197}$Au   as an example, where the Soft EOS with MDI has been used. Figure \ref{stability}
displays the binding energies and root mean square radii  of the cold projectile ($^{68}$Ni) and target ($^{197}$Au) nuclei 
as a function of time till 400 fm/c. It is clearly seen that both variables keep very stable during the whole time evolution, 
which assures that we can procced with the following calculations on the GDR and PDR.

\begin{figure}[htbp]
\resizebox{8.6cm}{!}{\includegraphics{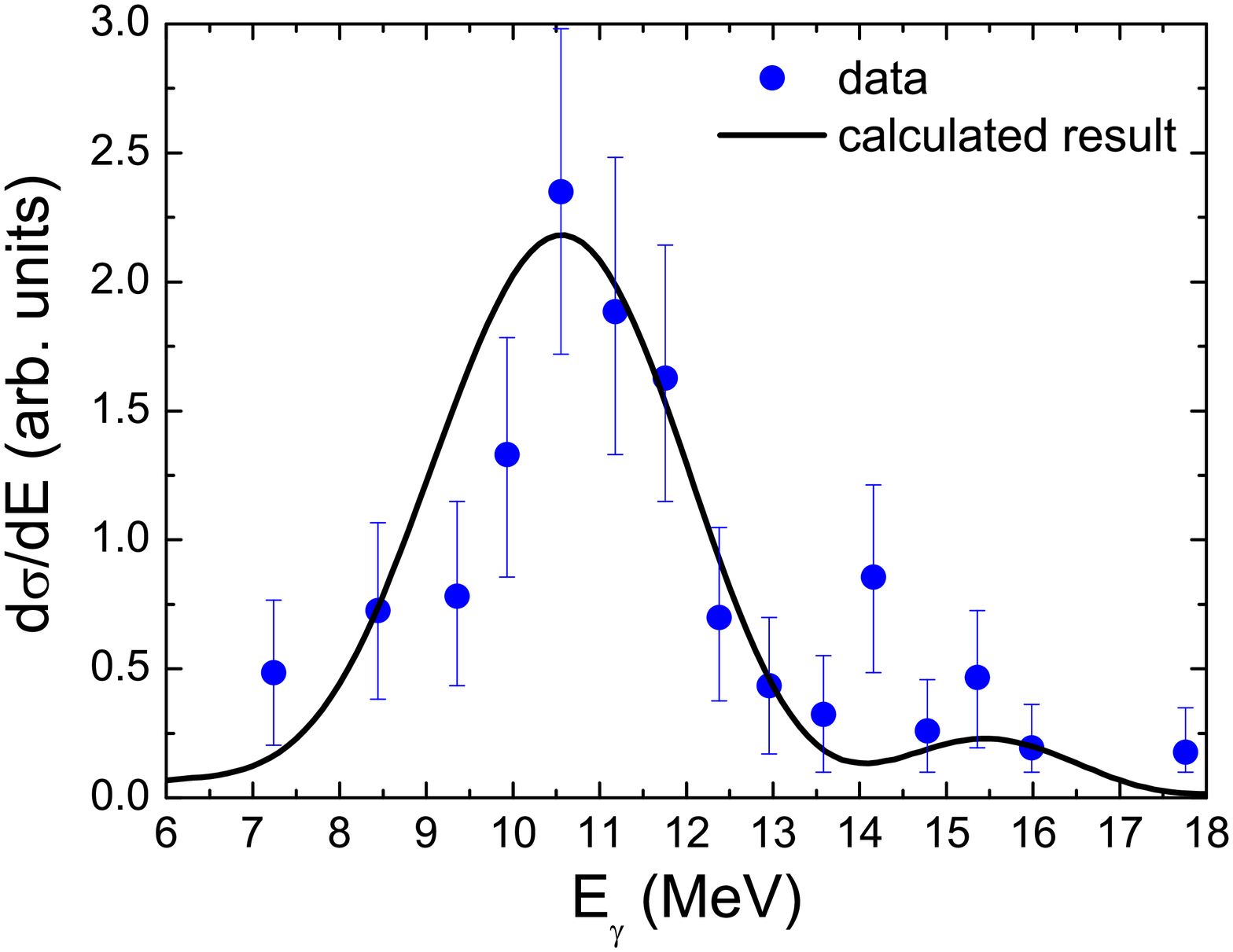}}
\caption{(Color online) The calculated PDR result of $^{68}$Ni
compared with  the experimental data. The blue circle with error
bar is the experimental data in Ref.~\cite{nidat}, the black line
is the calculated result.} \label{nidat}
\end{figure}

After the stability check, we shall compare the calculated
results with the experimental data to demonstrate the reliability
of our calculation.  The calculation parameters are as follows:
incident energy ($E_{in}$) is 600 MeV/nucleon, impact parameter
($b$) is 24fm, the Soft EOS with MDI, and  C$_{sym}$ is 32MeV. 
The 
PDR result of $^{68}$Ni is shown in Figure ~\ref{nidat}  where
the  symbol with error bar is for the experimental data from
Ref.~\cite{nidat}, the line is our calculated result. We can see
that the peak energy of our calculated result has good agreement
with the data even though the FWHM of our calculated result is a
little larger than the data. For GDR calculations, our previous
IQMD calculations have shown a good agreement with the data
\cite{Wu}. After this comparison, we perform a systematic
calculation to explore the variable dependencies of GDR
and PDR.

\begin{figure}[htbp]
\resizebox{8.6cm}{!}{\includegraphics{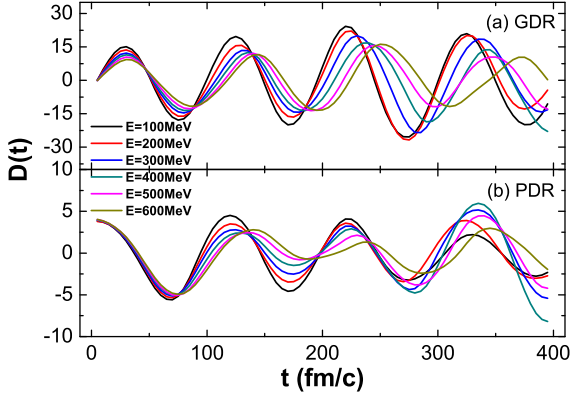}}
\vspace{-0.4cm} \caption{(Color online)  Time evolution of dipole
moment for GDR  (a) and PDR (b) at different incident energies. The
meaning of lines are illustrated in the insert. In calculations, we use $b$ = 24 fm,
C$_{sym}$ =32MeV, and the Soft EOS without MDI.  } \label{moment}
\end{figure}

\begin{figure}[htbp]
\resizebox{8.6cm}{!}{\includegraphics{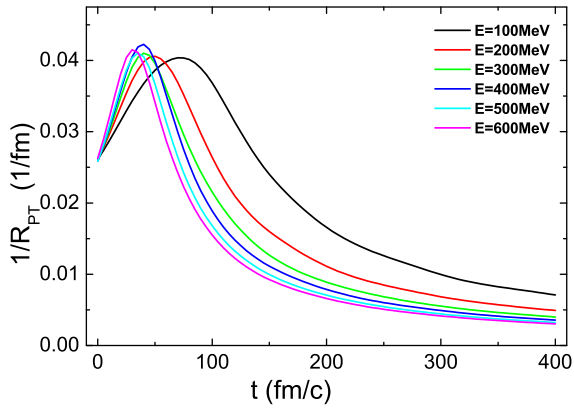}}
\vspace{-0.4cm} \caption{(Color online)  Time evolution of dipole
moment for GDR  and PDR at different incident energies. The
meaning of lines are illustrated in the insert.  In calculations, we use $b$ = 24 fm,
C$_{sym}$ =32MeV, and the Soft EOS without MDI.  } \label{dist}
\end{figure}

Firstly, the sensitivities of GDR and PDR of $^{68}$Ni in the
Coulomb excitation  of $^{197}$Au to some parameters have been
studied. From the calculated results, we can extract the peak
energy, strength and FWHM of GDR or PDR by Gaussian fitting to the
$\gamma$ emission spectrum. Figure~\ref{moment} shows the time
evolution of dipole moment for GDR and PDR at different incident
energies.  With an increasing of energy, both amplitude and
frequency of GDR and PDR oscillations become smaller, this could
be attributable to the  weakening of the effective Coulomb field in
which the projectile interacts with the target when the projectile has
faster velocity.  Actually this can be seen from the inverse of the distance ($1/R_{PT}$) between the CM of protons of projectile and CM
of protons of target. Figure ~\ref{dist} shows  $1/R_{PT}$ as a function of time in different incident energies.
If we roughly treat the projectile and target as point charged particles, then the Coulomb interaction is
proportional to the time  integration of  $1/R_{PT}$. Therefore, the Coulomb effect becomes smaller when
the incident energy is larger and then the projectile gets less excited.

\begin{figure}[htbp]
\resizebox{9.5cm}{!}{\includegraphics{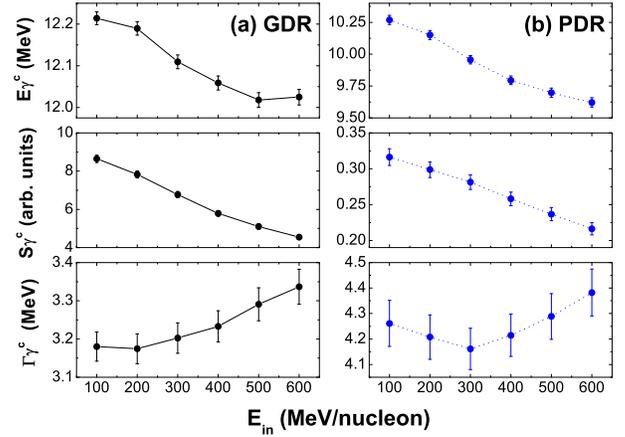}}
\vspace{-0.4cm} \caption{(Color online)Incident energy dependence
of dipole moments for $^{68}$Ni. The upper panel for GDR and the bottom
one for PDR.  In calculations, we use $b$ = 24 fm,  C$_{sym}$ =
32MeV, and the Soft EOS without MDI. \label{ve}}
\end{figure}

On the other hand,  in Figure~\ref{moment} pygmy dipole oscillation shows a much
smaller amplitude and a slightly smaller frequency in comparison
with the GDR case. These lead to the behavior of  the incident energy
dependence of the GDR and PDR parameters as shown in Figure~\ref{ve} .
With the increase of incident energy, both the peak energies and
strength for GDR and PDR  drop. For the FWHM, it clearly shows an
increasing behavior for GDR. Similarly,  FWHM of PDR displays an  increasing behavior
versus energy within large errors.

\begin{figure}[htbp]
\resizebox{9.5cm}{!}{\includegraphics{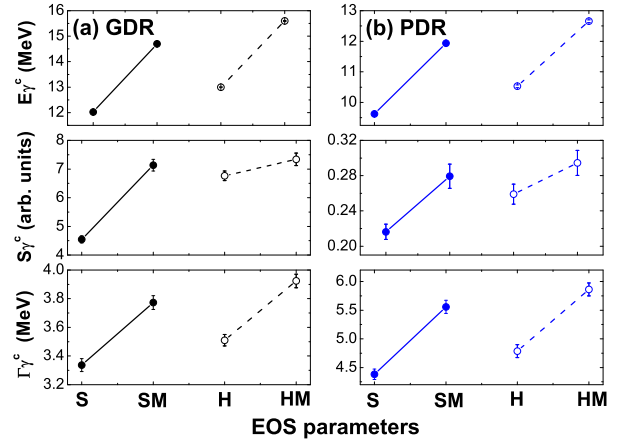}}
\vspace{-0.8cm} \caption{(Color online) EOS dependencies of GDR
(left panels) and PDR  (right panels) parameters for $^{68}$Ni. In calculations, we use
$E_{in}$= 600 MeV/nucleon,  $b$= 24 fm, and C$_{sym}$ = 32MeV. }
\label{eos}
\end{figure}

Figure~\ref{eos} shows EOS dependence of the dipole resonance
parameters. Compared to the Soft EOS with or without momentum
dependent interaction, the calculated results with the Hard EOS
are generally bigger. On the other hand, compared to the  EOS
without MDI, all the peak energy, strength and FWHM in the EOS
with MDI are also larger. In this sense, MDI has similar effect to
make the EOS harder.  From this figure, we learn that diploe
$\gamma$-emission is very sensitive to the stiffness of EOS.

\begin{figure}[htbp]
\resizebox{8.6cm}{!}{\includegraphics{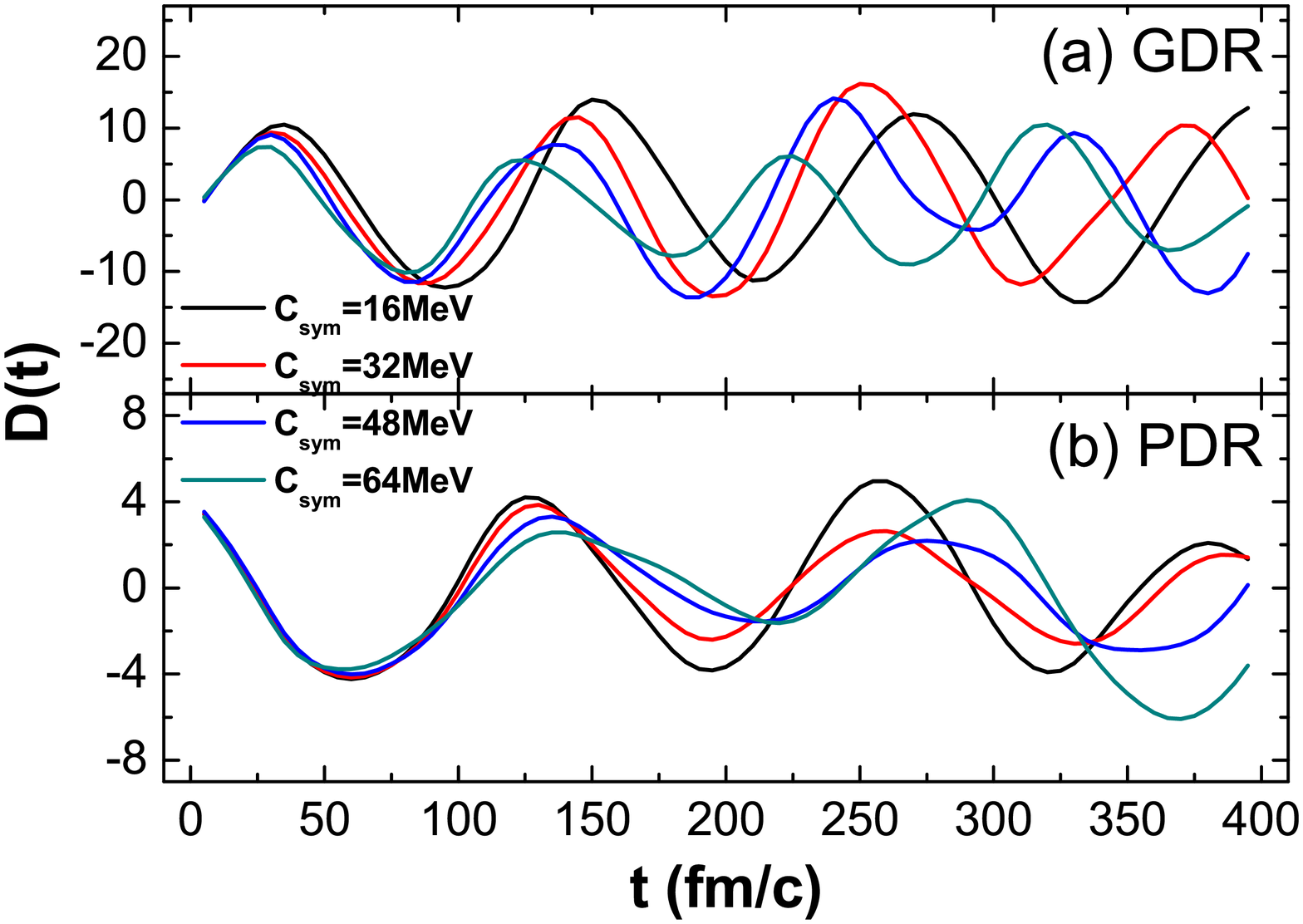}}
\vspace{-0.4cm} \caption{(Color online)C$_{sym}$ dependence of
of dipole moments for $^{68}$Ni.  The upper panel for GDR and the bottom
one for PDR.   In calculations, we use $E_{in}$= 600 MeV/nucleon, $b$ = 24 fm, and the Soft EOS without MDI. \label{csym0}}
\end{figure}

Many works have demonstrated that the symmetry energy
plays an important role in understanding the mechanisms of
many exotic phenomena in nuclear physics and astrophysics.
Dipole oscillations between all neutrons and all protons or
between the excess neutrons and a core
composed of an equal number of protons and neutrons
involve asymmetry of neutron and proton, which could be
influenced by the symmetry energy term of the EOS. To this end, we change the value of
C$_{sym}$ in IQMD model to address its effect on GDR and PDR. 
In order to see a clear trend of GDR and PDR toward the symmetry energy coefficient, we choose a larger range of 
C$_{sym}$ in the calculations, i.e. from 16 - 64 MeV. In between,  a value of C$_{sym}$ around 36 MeV 
has been thought reasonable to describe the property of ground state of nuclei in many previous studies. 
 Figure~\ref{csym0} shows time evolution of dipole moments
for GDR (upper panel) and PDR (lower panel)  in different symmetry
coefficients C$_{sym}$. It shows that the
frequency of GDR  oscillations becomes faster and the amplitude tends to be smaller  with the increasing of $C_{sym}$, which leads to an increasing of peak energy and decreasing of strength as shown in next figure.
On the contrary, the
frequency of PDR  oscillations becomes slightly slower, but the amplitude tends to be smaller as well as GDR  with the increasing of $C_{sym}$, which induces a decreasing of peak energy as well as strength as shown in next figure.

\begin{figure}[htbp]
\resizebox{9.5cm}{!}{\includegraphics{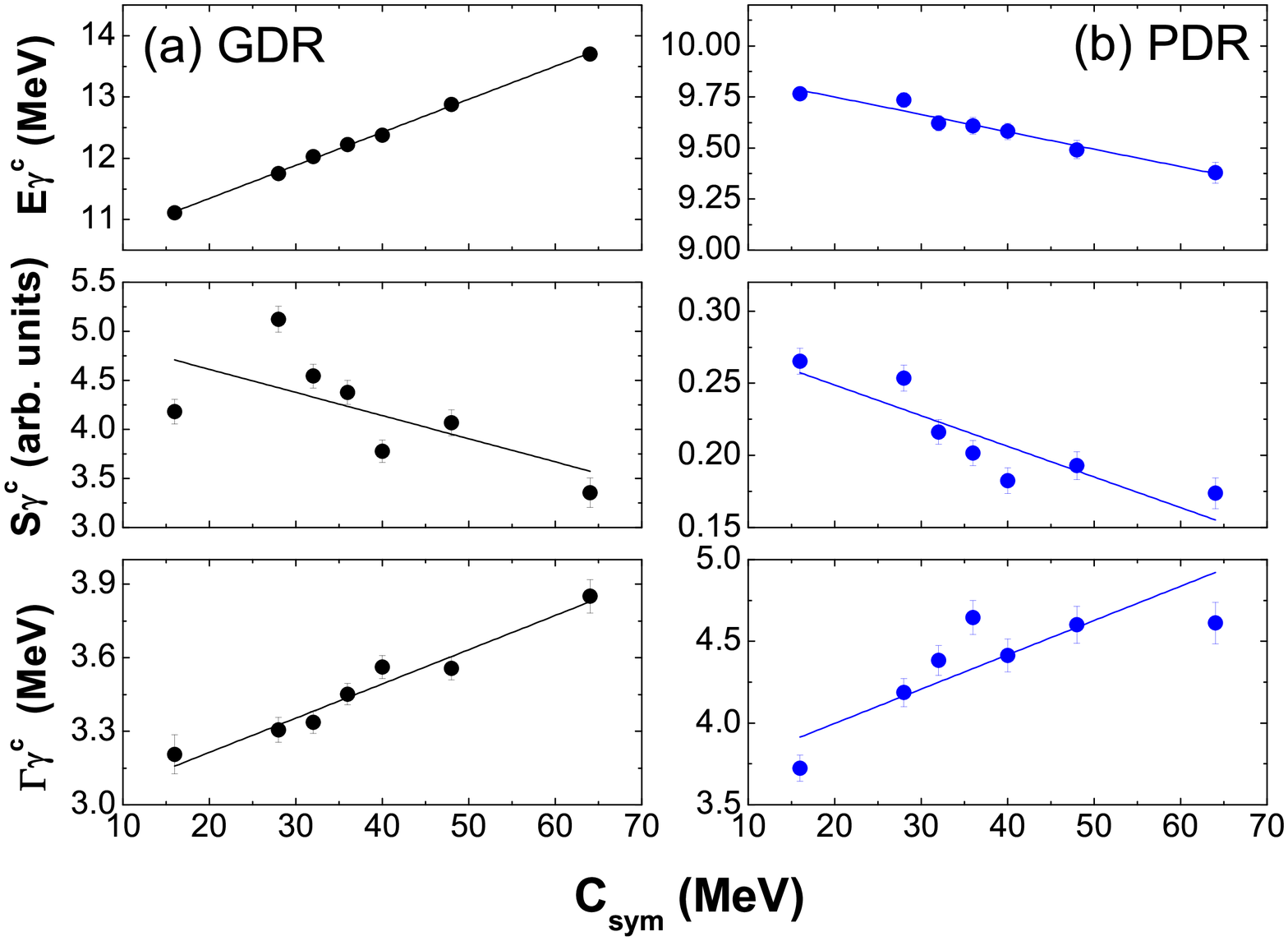}}
\vspace{-0.4cm} \caption{(Color online) C$_{sym}$ dependence of
GDR (left panels) and PDR  (right panels) parameters  for
$^{68}$Ni.  Symbols present for calculations and curves are linear fits  for guiding the eyes.  In
calculations, we use $E_{in}$= 600 MeV/nucleon, $b$ = 24 fm, and the Soft
EOS without MDI.  } \label{csym}
\end{figure}

Figure~\ref{csym} shows the calculated results of PDR and GDR parameters with the different  C$_{sym}$ parameters.  With the increases of C$_{sym}$, the peak energies of GDR  show
a linear increase, but  those of PDR shows a little decrease. However, considering the obvious different scale in y-axis for GDR and PDR,
peak energy of PDR has only about 4$\%$ decrease in contrast to 25$\%$ increase for GDR. For strength, both show a decreasing behavior and for FWHM, both display an increasing trend.
In some previous studies, it was  known that
the restoring force of isovector GDR is proportional to the
symmetry energy of nuclear matter, which makes the
peak energies of GDR pushed to higher energies with
stronger symmetry energy. By contrast,  PDR originates
mainly from the vibrations of a few valence neutrons which  against
the isospin neutral core. Those valence neutrons are located in the very low
density  exterior region, where the symmetry energy is not affected as much.
In general, the main difference by $C_{sym}$ is observed
in the interior region where is responsible for GDR, which has very little or weak influence on the
subtle properties of PDR. Similar insensitivity of PDR to symmetry energy was also
found in some literatures \cite{MaZY,Baran}.

\begin{figure}[htbp]
\resizebox{9.5cm}{!}{\includegraphics{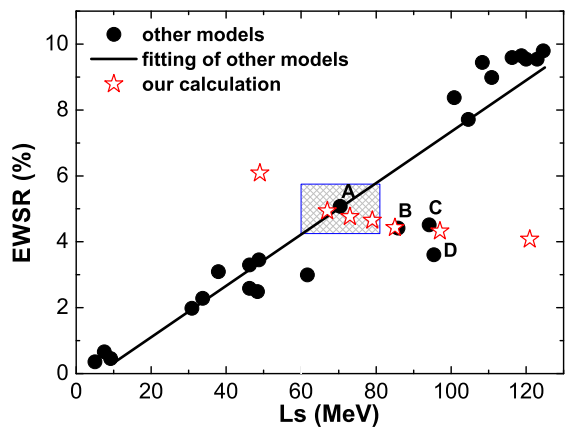}}
\caption{(Color online) Calculated  EWSR ($PDR_{m_1}/(PDR_{m_1}+GDR_{m_1})\%$) for $^{68}$Ni with different $E_{sym}$ in our IQMD model (open stars) together with a bunch of  mean-field model calculations (solid circles), which is taken from Ref.~\cite{Ita}.  Line is a fit to  mean-field model calculations. In the figure, A, B, C and D represent mean-field calculations with SkMP, SkRs, SkGs, and SK255 parametrization. For details, see Ref.~\cite{Ita}. In figure, the shaded area represents 15$\%$ uncertainty around the point A which is just on fitted line.
}
\label{EWSR}
\end{figure}

It will be very interesting if we can pin down some information on the symmetry energy, such as
the $E_{sym}(\rho_0)$ and the derivative of the symmetry energy at
saturation ($L_s$) by the dipole resonance calculation. In our IQMD model, the symmetry energy reads
\begin{equation}
E_{sym} =  12.5 (\frac{\rho}{\rho_0})^{2/3} + \frac{C_{sym}}{2} (\frac{\rho}{\rho_0}),
\end{equation}
where the first term  is the kinetic energy contribution and the second term is potential energy term.
In this case, the slope parameter of the symmetry energy at saturation, $L_s = 3 C_{sym}/2 + 25$.  With the various $C_{sym}$, we can obtain a plot of EWSR $PDR_{m_1}/ (PDR_{m_1}+GDR_{m_1})\%$ versus $L_s$ as shown in Figure \ref{EWSR}.
To compare with other model calculations, different mean-filed calculation results  \cite{Ita} for $^{68}$Ni are displayed in the figure.  From the figure, we observe a weak anticorrelation of EWSR vs $L_s$ in IQMD calculations. However, the IQMD results are quite similar to the mean-filed calculations with SkMP, SkRs, SkGs, and SK255 parameterizations, i.e. the corresponding A, B, C and D points \cite{Ita}. The anticorrelation is
not so reasonable in comparison with the overall positive correlation trend which has been constructed by the mean-field calculations from $L_s$ $\approx$ 10-120 MeV.  This may originate from the unreal symmetry energy coefficient  when $C_{sym}$ is selected too large or too smaller.  Assuming the fitting line is correct for the relationship of EWSR versus $L_s$ and allow 15$\%$ uncertainty, we can obtain that the reasonable $L_s$ values from 60 to 81 MeV for IQMD calculations (see the shaded area), which leads to $C_{sym} \approx$ 23.3 - 37.3 MeV ,
and gives $E_{sym}(\rho_0) \approx$   24.2 - 31.2 MeV.  The above values for IQMD are in  good agreement with other calculations.  For instance,
in Ref.~\cite{Ita}, they got the weighted average, $L_s$ = 64.8 $\pm$ 15.7 MeV, and the deduced best value of $E_{sym}(\rho_0)$ = 32.3 $\pm$ 1.3MeV. In  Ref.~\cite{Kli}
they give the value of $E_{sym}(\rho_0)$ = 32.0 $\pm$ 1.8 MeV  and in Ref.~\cite{Tsang}
the ranges obtained of $E_{sym}(\rho_0)$ is 30.2 - 33.8MeV. So far, the above $E_{sym}(\rho_0)$ and $L$ parameters give a useful constraint for the IQMD model.

\begin{figure}[htbp]
\vspace{-1.cm}
\resizebox{10.cm}{!}{\includegraphics{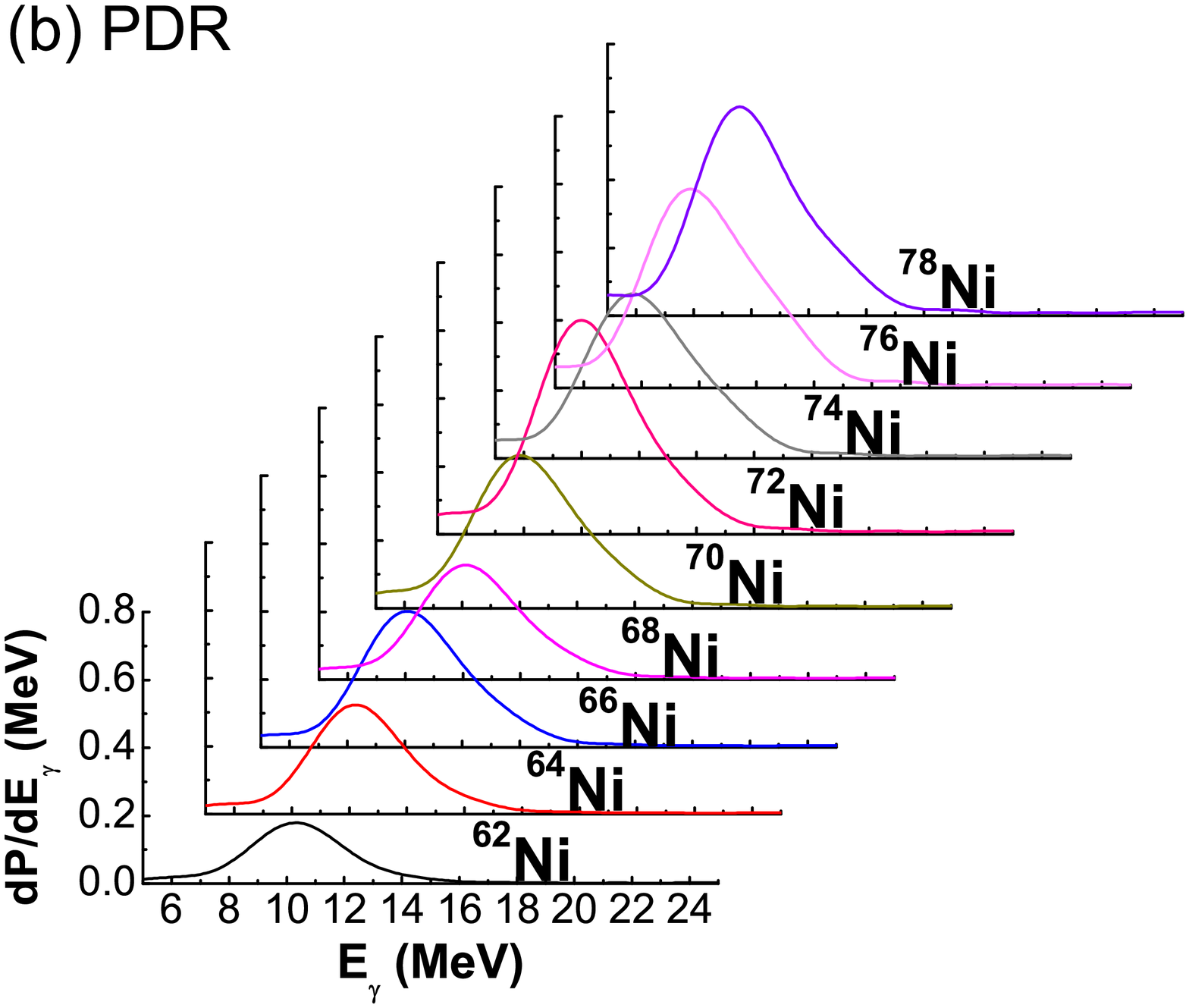}}
\vspace{-0.4cm} \caption{(Color online) Mass number dependence of GDR (a) and PDR spectra (b)  for
Ni isotopes. In calculations, we use $E_{in}$= 100 MeV/nucleon, $b$= 24fm,
C$_{sym}$ =32 MeV, and the Soft EOS without MDI.  } \label{mom_A}
\end{figure}

We also studied  the systematic evolution of GDR and PDR
parameters of even-even Ni  isotopes from $^{62}$Ni to $^{78}$Ni.
Figure \ref{mom_A} shows the distributions of isovector dipole strength for
the Ni isotopes.  In the figure, the GDR peak is
around 12 MeV and that of PDR is around 10 MeV. The
PDR strengths increase as the neutron number increases.

\begin{figure}[htbp]
\resizebox{9.5cm}{!}{\includegraphics{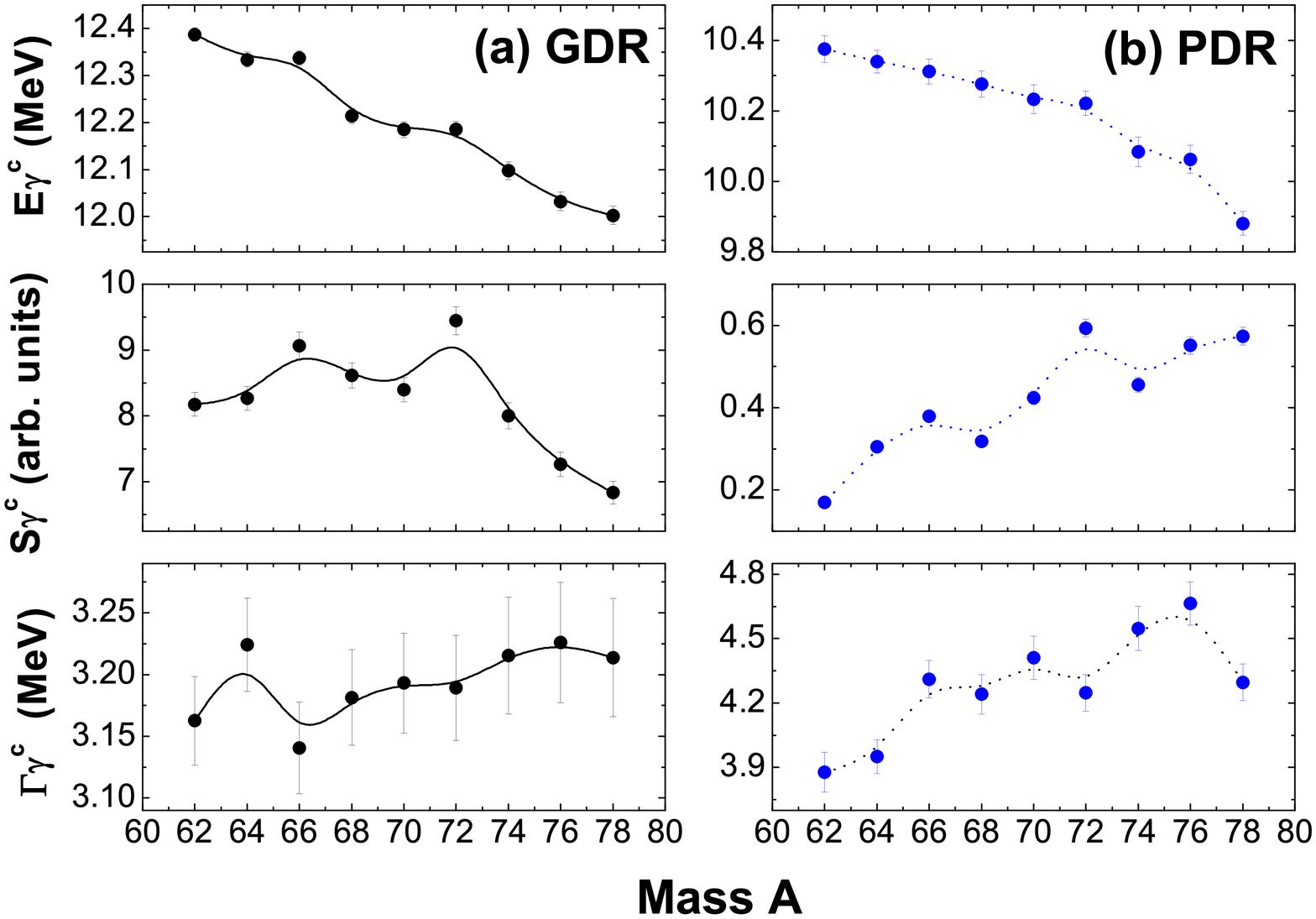}}
\vspace{-0.4cm} \caption{(Color online) Mass number dependence of
Ni isotopes of GDR  (left panels) and PDR (right panels)
parameters. In calculations, we use $E_{in}$= 100 MeV/nucleon, $b$= 24fm,
C$_{sym}$ =32 MeV, and the Soft EOS without MDI. } \label{nix}
\end{figure}

Figure~\ref{nix} shows the extracted  GDR and PDR parameters  for Ni isotopes from Figure~\ref{mom_A}.
With the increases of mass number $A$ or neutron excess, the peak
energies of both GDR and PDR  decrease. This behavior reflects
that a larger neutron excess results in lower GDR and  PDR
excitation energy. In the region beyond $A$ = 72, the slope of the
PDR peak energies becomes steeper than those of lower mass nuclei,
because the neutrons in outer orbital are more loosely bound and
thus the restoring force in the oscillation of the skin against
the core becomes weaker. The strength of GDR is
shown to seemingly decrease
decreasing for larger mass nuclei but it increases for PDR, the
FWHM of both GDR and PDR show increasing behavior with $A$, except
for smaller $A$ for GDR.

\begin{figure}[htbp]
\resizebox{10.cm}{!}{\includegraphics{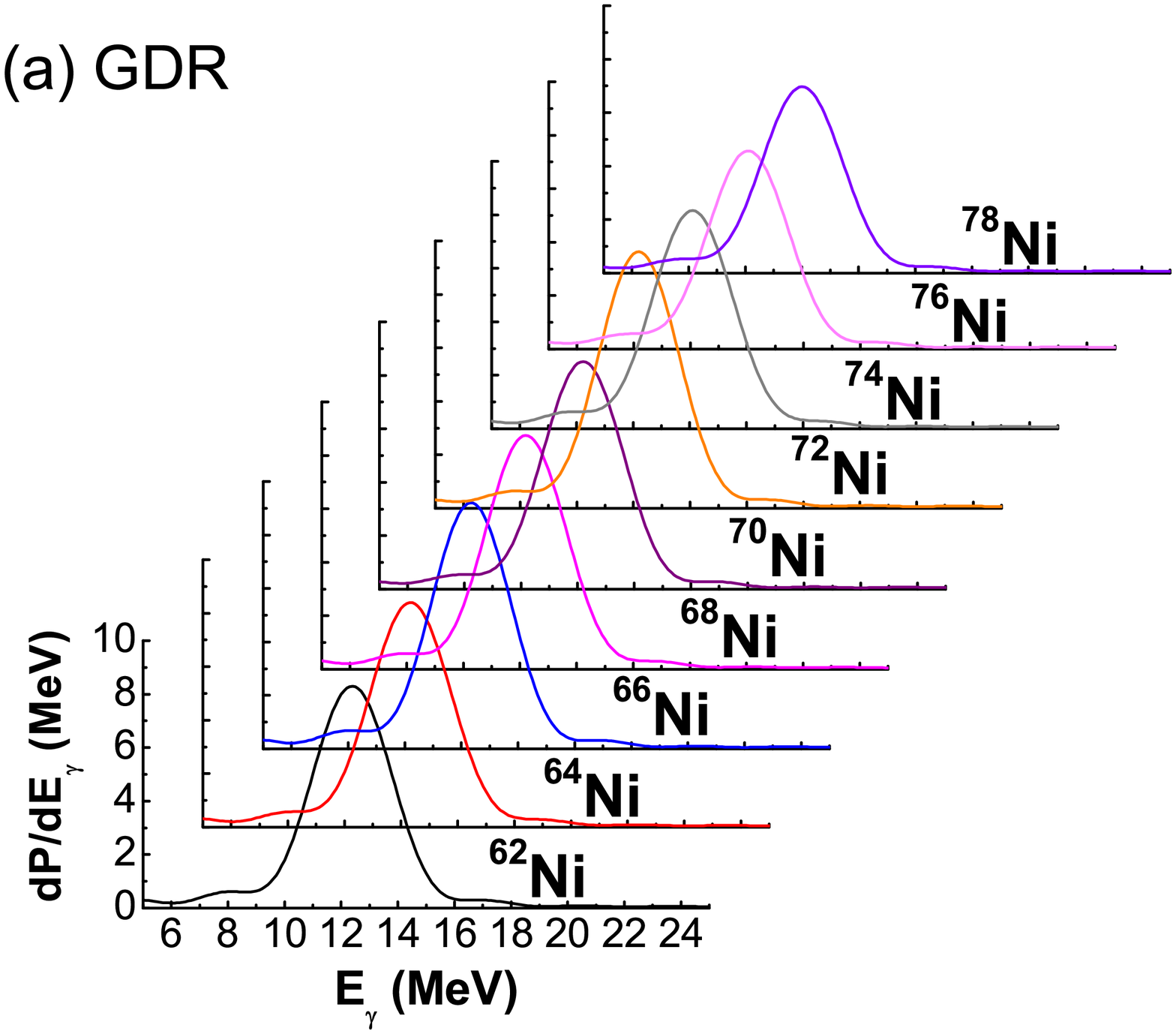}}
\end{figure}

\begin{figure}[htbp]
\resizebox{9.5cm}{!}{\includegraphics{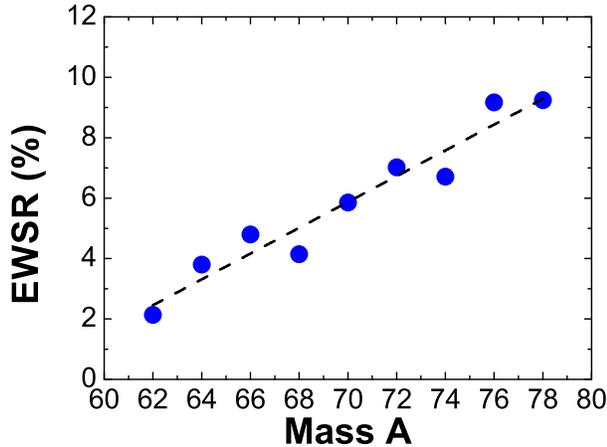}}
\caption{(Color online) Calculated  EWSR $PDR_{m_1}/GDR_{m_1}\%$ for Ni isotopes (solid circles) together with a linear fit (line).
}
\label{pvg}
\end{figure}

Finally, the fraction of the EWSR  $m_1 \%$ is also extracted. Figure~\ref{pvg} shows the calculated results which clearly illustrate that with the increases of mass number $A$ or neutron-skin thickness, PDR component increases almost linearly. Similar phenomenon  has been also reported earlier for some  isotopes with random phase approximation phenomenological approach \cite{PDRCAL6,Co,Jap,Ita}.

\section{Summary}

In summary,  we have applied the IQMD model to study Giant and
Pygmy Dipole Resonance in Ni isotopes by the Coulomb excitation.
Similar to  the method to calculate Giant Dipole Resonance in our previous IQMD calculation, we
extend it to calculate PDR in the same model. We first showed very stable initial projectile and target can be obtained with the soft momenteum-dependent EOS, which assures  the performance of GDR and RDR calculations. After we got a satisfied calculated PDR result for 
$^{68}$Ni  which agrees with the experimental data very well, then we
performed a systematic calculation for GDR and PDR with different
beam energy, EOS and symmetry energy parameters for
$^{68}$Ni+$^{197}$Au situation. It is found that the peak energy and strength for both GDR and PDR show
a decreasing trend with the incident energy, which can be understood by the excitation extent due to Coulomb field in different energies. When the equation of state becomes harder or momentum dependent interaction is taken into account, the increasing behaviors of peak energy, strength and FWHM emerge for both GDR and PDR.  Concerning the symmetry term of EOS, it shows strong positive correlation with the GDR's peak energy and FWHM, but it plays a relative weak role on PDR.  By fitting to other mean-field calculations of EWSR vs the slope parameter at saturation $L_s$ and assuming 15$\%$ uncertainty of $L_s$, i.e. $L_s$ = 60 -81 MeV, we can obtain the range of symmetry energy coefficient $C_{sym} \approx$ 23.3 - 37.3 MeV and  symmetry energy  at saturation is around   24.2 - 31.2 MeV and  in the IQMD model, which gives a constraint for the QMD model.
 Finally, mass number or isospin dependence of
GDR and  PDR for the Ni isotopes is presented. The result shows that
the peak energy of both GDR and PDR decreases with mass number, and the FWHM of them increases. For the strength,  PDR shows an increasing behavior but GDR seemingly decrease with mass of isotopes. For the fraction of the EWSR in
Ni isotopes, it shows a linear rise-up with the increases of mass
number $A$ or neutron excess.

\section*{Acknowledgements}

This work was supported in part by  the
National Natural Science Foundation of China under Contract Nos.
11035009, 10979074, and 11205230, the Major State Basic Research
Development  Program in China under Contract No. 2013CB834405, the Chinese Academy of Science Foundation
under Grant No. KJCX2-EW-N01.


\end{document}